\documentclass[pra,superscriptaddress,10pt]{revtex4}

\usepackage[textures]{graphics}

\newcommand{\beq}{\begin{equation}}
\newcommand{\eeq}{\end{equation}}
\newcommand{\beqa}{\begin{eqnarray}}
\newcommand{\eeqa}{\end{eqnarray}}
\newcommand{\ket} [1] {\vert #1 \rangle}
\newcommand{\bra} [1] {\langle #1 \vert}
\newcommand{\braket}[2]{\langle #1 | #2 \rangle}

\begin{document}
\title{About economic qubit cloning}
\author{Thomas Durt}
\affiliation{TONA-TENA Free University of
Brussels, Pleinlaan 2, B-1050 Brussels, Belgium}
\author{Jiangfeng Du}
\affiliation{Structure Research Laboratory and Department of Modern Physics, University of
Science and Technology of China, Hefei, Anhui 230026, China}
\affiliation{Department of Physics, National University of Singapore, Lower Kent Ridge,
Singapore 119260, Singapore}

\begin{abstract}
In this paper we establish a deep connection between the 3 qubit
one-to-two phase-covariant quantum cloning network of Fuchs et al. [C.
Fuchs, N. Gisin, R.B. Griffiths, C.S. Niu, and A. Peres, Phys. Rev. A {\bf 56}
$n^{\circ}$4, 1163 (1997)], and its economic 2 qubit counterpart due to
Niu and Griffiths [Phys.Rev. A {\bf 60}
$n^{\circ}$4, 2764 (1999)]. A general, necessary and sufficient criterion
is derived in order to characterize the
reducibility of 3 qubit cloners to 2 qubit cloners. When this criterion is
fulfilled, economic cloning is possible. We
show that the optimal isotropic or universal 3 qubit cloning machine is not
reducible to a 2 qubit cloner.
\end{abstract}


\maketitle

\section{Introduction}

Quantum Information emerged from the fruitful cross-fertilization of
quantum
mechanics and information technology. In the last decade, particularly
promising applications such
as quantum cryptography, quantum cloning, quantum teleportation, quantum
games and quantum computers
were implemented experimentally, with more or less succes
\cite{RMP,Cummins,tele,du,Gulde}.
Although it is not certain that
these progresses will lead to a practical quantum computer \cite{Jones},
because
of the difficulties inherent to decoherence, quantum cryptography is
already a grown up, user
friendly and efficient technology \cite{RMP,naturecerf}.
Traditionally, it is
implemented with two-level quantum systems, known as qubits. The
inviolability of the quantum key
distribution protocols such as the BB84 protocol \cite{BB84} is guaranteed
by
the {\it no-cloning}
theorem \cite{Dieks,Wootters} that states that perfect copying
(cloning) of a set of input
states that contains at least two non-orthogonal states is impossible. It
is however possible to
realize approximate quantum cloning, a concept that was introduced in a
seminal paper of Buzek and
Hillery \cite{buzek}, where a universal (or state-independent) and
symmetric
one-to-two cloning
transformation was introduced for qubits. Here we are rather interested
in
state-dependent, symmetric or asymmetric
one-to-two quantum cloning, in particular in the so-called
phase-covariant
qubit transformation \cite{FGGNP,bruss,NG}
that optimally copies (with the same fidelity) all the pure states of the
form ${1\over
\sqrt 2}(\ket{0}^Z+e^{i\phi}\ket{1}^Z)$ (where $\phi \in [0,2\pi]$ while
$\ket{0}^Z$ and $\ket{1}^Z$
represent up and down spin states along a conventional direction $Z$).
Such
``equatorial'' states
are located on the intersection of the $XY$ plane with the Bloch sphere.
It
is important in the
context of quantum cryptography to study the performances of such cloners
because they condition the
security of quantum cryptographic protocols. For instance the optimal
phase-covariant cloner is
presently believed to provide the most dangerous eavesdropping strategy
for
the BB84 quantum
cryptographic protocol \cite{BB84}. This justifies the interest of implementing
experimentally the qubit
one-to-two phase covariant cloners that were theoretically proposed in
the
past. The first
theoretic proposal, that we shall from now on denote the FGGNP proposal,
required
the
use of 3 qubits\cite{FGGNP}, two in addition to the one
carrying
the original signal.
Unfortunately, recent NMR experiments that were realized in order to
implement the (3 qubit)
universal qubit cloner \cite{Cummins} showed that a substantial loss
occured, due to
inhomogeneities of the magnetic field and decoherence. It was difficult
during that experiment to
avoid such effects because the corresponding cloning network contained no
less than ten single qubit
gates and five 2 qubit gates. Moreover, it required to control with enough
precision the two-by-two
entanglement between three qubits which is not a simple task at all. A
quick analysis of the
results of \cite{FGGNP} shows that an experimental implementation of their
proposal for phase-covariant cloning would
certainly face the same problems. Fortunately, in the case of
phase-covariant cloning,
there exists in theory a simplified, ``cheap'' or ``economic'' 2 qubit
network due to Niu and
Griffiths \cite{NG}, that we shall from now on denote the NG proposal, in
which no external ancilla
is required and that exhibits the same cloning properties as the FGGNP
proposal. This proposal is
quite simpler to implement experimentally, it requires only two single
qubit
gates and one 2 qubit
gate, and it requires to control the entanglement of a pair of qubits
only. It is thus likely to
be quite less noisy than its 3 qubit counterpart. The goal of the present
paper is to elucidate the connections that exist
between 2 and 3 qubit cloners. It can be seen as a first step towards a
study of the possibility to replace 3 qu$N$it cloners
by (economic) 2 qu$N$it cloners, a problem that presents a real interest in
connection with the security of quantum
cryptographic protocols.

In the second section, we introduce some useful formal tools and establish
a deep relation between the FGGNP
transformation and
the NG transformation, in the sense that we prove that the FGGNP
transformation is
equivalent to a symmetrised
version of the NG transformation. In the third section, we derive a general
theorem that defines precisely under which
conditions a 3 qubit cloner is reducible to a 2 qubit cloner. It allows us
to prove the non-existence of  such a
relation in the case of
universal cloning.

 A
summary of our results, conclusions, and a brief discussion of some open
problems are presented in the
last section.

\section{3 qubit and 2 qubit phase-covariant cloners, and
connections
between them.}

\subsection{(A strictly covariant generalisation of) Cerf's formalism for
3 qubit phase-covariant cloners}
Before we introduce Cerf's formalism for cloning machines it is useful to
recall the properties of the so-called
Bell states. The four Bell states are defined as follows:

\beq\label{Bell}\ket{B^{Z}_{m,n}}_{1,2}={1\over\sqrt{2}}\sum_{k=0}^{1}
(-)^{k.n}\ket{k}^Z_{1}\ket{k+m}^Z_{2}\eeq where $m,n \in \{0,1\}.$

Consequently: \beqa
&&\ket{B^Z_{0,0}}_{1,2}={1\over \sqrt
2}\{\ket{0}^Z_{1}\ket{0}^Z_{2}+\ket{1}^Z_{1}\ket{1}^Z_{2}\} ,\qquad
\ket{B^Z_{0,1}}={1\over \sqrt
2}\{\ket{0}^Z_{1}\ket{0}^Z_{2}-\ket{1}^Z_{1}\ket{1}^Z_{2}\}\\
&&\ket{B^Z_{1,0}} ={1\over \sqrt
2}\{\ket{0}^Z_{1}\ket{1}^Z_{2}+\ket{1}^Z_{1}\ket{0}^Z_{2}\},\qquad
\ket{B^Z_{1,1}}={1\over \sqrt
2}\{\ket{0}^Z_{1}\ket{1}^Z_{2}-\ket{1}^Z_{1}\ket{0}^Z_{2}\}
\eeqa
 where
$\ket{0(1)}^Z_{1(2)}$ represents a spin up
(down) state of the qubit system $1$ ($2$) along a conventional direction
(here the
$Z$ direction). They are maximally entangled states and form an
orthonormal
basis of
the 4-dimensional Hilbert space spanned by the qubit states $1$ and $2$.

Let us now consider the following situation: Alice sends to Bob qubits
that
are either spin up or spin
down along $Z$ with 50-50 probability. Note that this is equivalent to a
situation during which Alice and Bob
share the maximally entangled state
$\ket{B_{0,0}}_{A,B}$, while Alice measures the spin of her qubit along
$Z$. N. Cerf proposed in
Refs.~\cite{CERFPRL,CERF} a general characterization of asymmetric and
state-dependent
$1\to 2$ cloning transformations for $2$-level systems which is, roughly
speaking, summarised as follows. Eve
copies the state $\ket{B_{0,0}}^Z_{A,B}$ by replacing it by the cloning
state, which is assumed to be a 4 qubit
state (one qubit for Alice, one for Bob, one for Eve and one ancilla). We
shall
from now on denote it
$\ket{\Psi}^Z_{A,B,E,M}$ where the indices are representative of the
reference qubit possessed by Alice ($A$), of
the two output clones ($B$ for Bob and
$E$ for Eve), and of the ($2$-dimensional) ancilla or cloning machine
($M$). According to Cerf's ansatz, the
cloning state is biorthogonal in the Bell bases, which imposes that
\beq \label{ansatz}\ket{\Psi}^Z_{A,B,E,M}=\sum_{m,n=0}^1
a_{m,n}\ket{B_{m,n}}^Z_{A,B}\ket{B_{m,n}}^Z_{E,M}\eeq where $a_{m,n}$ is a
(normalised) 2x2 matrix. The specification of the amplitudes $a_{m,n}$
defines the cloning
transformation. Remark that such a state can be obtained by letting work
on
the initial state
$\ket{B_{0,0}}^Z_{A,B}\ket{B_{0,0}}^Z_{E,M}$ a unitary transformation of
the type $1_A\otimes U_{BE}\otimes 1_M$
that affects neither Alice's qubit, nor the ancilla (although the
preparation of the initial state requires that Eve
entangles the clone and the ancilla).

The deep reason
therefore is that Bell states $\ket{B_{m,n}}^Z_{1,2}$ can be generated
from
an initial state prepared along
$\ket{B_{0,0}}^Z_{1,2}$ via local transformations. For instance, we have
$1_1\otimes\sigma_2^{X}
\ket{B_{0,0}}^Z_{1,2}=\ket{B_{1,0}}^Z_{1,2}$, $1_1\otimes\sigma_2^{Y}
\ket{B_{0,0}}^Z_{1,2}=i\ket{B_{1,1}}^Z_{1,2}$, $1_1\otimes\sigma_2^{Z}
\ket{B_{0,0}}^Z_{1,2}=\ket{B_{0,1}}^Z_{1,2}$ where the $\sigma$'s are the
Pauli matrices. In virtue of this
property, it is not absolutely
necessary that Alice and Bob share a maximally entangled state to begin
with: we can as well consider the situation in which Alice sends
directly a qubit to Bob. Nevertheless, it is convenient to
consider directly the cloning state in a 4 qubit space because as we
shall
see now there exists a covariant generalization of Cerf's formalism in
which a ``mirror''
relation
exists between Alice's qubit and the
cloning machine at one side, and between Bob's clone and Eve's clone at
the
other side.

Before describing this
generalization, it is useful to introduce and to motivate the concept of
strict covariance. In order to do so, let
us consider the $Z'$ basis defined as follows: $\ket{0}^{Z'}=\ket{0}^Z$,
$\ket{1}^{Z'}=i\ket{1}^{Z}$. Obviously
the following identities are satisfied:
$\ket{B_{0,0}}^{Z'}_{1,2}=\ket{B_{0,1}}^Z_{1,2}$,
$\ket{B_{0,1}}^{Z'}_{1,2}=\ket{B_{0,0}}^Z_{1,2}$,
$\ket{B_{1,0}}^{Z'}_{1,2}=i \ket{B_{1,0}}^Z_{1,2}$, and
$\ket{B_{1,1}}^{Z'}_{1,2}=i \ket{B_{1,1}}^Z_{1,2}$. Because of this, the
amplitudes of the cloning state
$\ket{\Psi}^Z_{A,B,E,M}=\sum_{m,n=0}^1
a_{m,n}\ket{B_{m,n}}^Z_{A,B}\ket{B_{m,n}}^Z_{E,M}$ are not necessarily
the
same in the primed basis. For instance
$\bra{0}^{Z'}_{A}\bra{1}^{Z'}_{B}\bra{0}^{Z'}_{E}\bra{1}^{Z'}_{M}\ket{\Psi}
^Z_{A
,B,E,M}=
(-)\bra{0}^{Z}_{A}\bra{1}^{Z}_{B}\bra{0}^{Z}_{E}\bra{1}^{Z}_{M}\ket{\Psi}^Z_{
A,B
,E,M}$.
We shall say that $\ket{\Psi}^Z_{A,B,E,M}$ is not strictly covariant when
we pass from the $Z$ to the $Z'$ basis.
Nevertheless, all the expectation values, of the type
$\bra{ijkl}^{Z(Z')}_{A,B,E,M}\ket{\Psi}^Z_{A,B,E,M}\bra{\Psi}^Z_{A,B,E,M}\ket{
ij
kl}^{Z(Z')}_{A,B,E,M}$,  which
are the diagonal coefficients of the density matrix considered in the
product basis assigned to the local detectors
labelled by $A,B,E,M$, are the same in both bases. As these are the
single
quantities that are of physical interest
in the present context, we can say that although the state
$\ket{\Psi}^Z_{A,B,E,M}$ is not strictly covariant when
we pass from the $Z$ to the $Z'$ basis it is covariant FAPP.

Now, let us recall that the theory of cloning machines was developed for
estimating the safety of quantum
cryptographic protocols. In such protocols, the information is encoded in
at least two non-commuting bases, which
guarantees that a perfect cloning process is impossible, in virtue of the
no-cloning theorem. Because of
this limitation, the best that Eve can do for eavesdropping the signal is
approximate cloning, and optimal
approximate cloning corresponds, as far as we know, to the most dangerous
(unperfect)
eavesdropping strategy that Eve is able to resort
to. It is usually assumed, conservatively, that Eve has perfect
technology
at her disposal (perfect transmission
lines, perfect quantum devices and so on), and that she dissimulates the
presence of her unperfect cloner under
the noise that would be otherwise attributed to an unperfect transmission
lines. Usually, transmission lines are
isotropic, which implies that the error rate or transmission noise is the
same in all encryption bases so that the
cloners must fulfill a fundamental constraint: they must exhibit the same
fidelities in all encryption bases
(the fidelity $F$ is defined for a dichotomic signal as follows: $F$=1-$e$
where $e$ is the
error rate). In the following, we shall
be interested in cloners that (i) satisfy the Cerf ansatz, and (ii) that are
strictly covariant when we pass from one
encryption basis to the other. This constraint is certainly exagerated
because in principle FAPP covariance (even
less, FAPP covariance in Alice and Bob's encryption/decryption bases
only)
is a sufficient constraint in order
that the cloner provides an unthwartable eavesdropping strategy.
Nevertheless,
it can be checked that all the
interesting cloners in the litterature, without exception, satisfy the
conditions (i) and (ii). It is not our goal in the
present paper to motivate why these conditions seem to be so natural in
the
study of cloning machines (see for
instance the fourth section of Ref.\cite{kwek} for a tentative explanation
of the generality of
Cerf's ansatz.). We shall show in the rest
of this section that once they are accepted, it is very easy to
establish a
deep connection between FGGNP's 3 qubit phase-covariant cloner and NG's 2
qubit phase covariant cloner.

Let us come back to the $Z$-$Z'$ transformation previously introduced.  As
we noted already,
the first
Bell state  $\ket{B_{0,0}}$ is not invariant:
$\ket{B_{0,1}}^Z_{1,2}=\ket{B_{0,0}}^{Z'}_{1,2}\not=
\ket{B_{0,0}}^Z_{1,2}=\ket{B_{0,1}}^{Z'}_{1,2}$. There is
another way to show this dissymetry between the $Z$ and the $Z'$ bases:
formally, if Alice and Bob share the
maximally entangled state $\ket{B_{0,0}}^Z_{A,B}$, and that Alice wants
to
transmit to Bob states that are
encrypted in the primed basis, she must encode her own qubits into the
conjugate basis $Z'^*$ defined as follows:
$\ket{0^*}^{Z'}=\ket{0}^Z$, $\ket{1^*}^{Z'}=-i\ket{1}^{Z}$. Indeed, it is
easy to check the identity
\beq\ket{B_{0,0}}^Z_{A,B}={1\over \sqrt
2}\{\ket{0}^Z_{A}\ket{0}^Z_{B}+\ket{1}^Z_{A}\ket{1}^Z_{B}\}={1\over \sqrt
2}\{\ket{0^*}^{Z'}_{A}\ket{0}^Z_{B}+\ket{1^*}^{Z'}\ket{1}^Z_{B}\}\eeq This
identity is the special case of a very
general property: let us consider an arbitrary basis
$(\ket{\psi_0},\ket{\psi_1})$ that we denote the
$\psi$ basis (with
$\braket{i}{\psi_j}$ = $U_{ ij}$). If Alice and Bob share the joint
state
$\ket{B_{0,0}}^Z_{A,B}$ and that Alice wishes to encode the signal in the
$\psi$ basis, she must project her
component of $\ket{B_{0,0}}^Z_{A,B}$ into the (conjugate) $\psi^*$ basis
defined as follows:
$\braket{i}{\psi^*_j}$ = $U^*_{ ij}$. This property is obvious if we note
that, in virtue of the unitarity
of $U_{ ij}$,
\beqa\ket{B_{0,0}}^Z_{A,B}={1\over \sqrt
2}\{\ket{0}^Z_{A}\ket{0}^Z_{B}+\ket{1}^Z_{A}\ket{1}^Z_{B}\} =
{1\over\sqrt{2}} \sum_{k,l,m=0}^{1}\ket{\psi^*_l}_{A}
\braket{\psi^*_l}{k}_{A}\otimes\ket{\psi_m}_{B}\braket{\psi_m}{k}_{B}
\nonumber\\= {1\over\sqrt{2}}
\sum_{k,l,m=0}^{1}
\ket{\psi^*_l}_{A}U_{ kl}\ket{\psi_m}_{B}U^*_{ km} = {1\over\sqrt{2}}
\sum_{k,l,m=0}^{1}
\ket{\psi^*_l}_{A}\ket{\psi_m}_{B}\delta_{ ml} =
{1\over\sqrt{2}} \sum_{k=0}^{1}
\ket{\psi^*_k}_{A}\ket{\psi_k}_{B}\label{identite}\eeqa
It would
be nice to modify slightly Cerf's
ansatz in order that this basic invariance property is respected. As it
was
shown by one of us \cite{kwek,zuko,nagler},
it is possible to achieve this goal thanks to a redefinition of Bell
states. These generalized Bell
states can be shown to obey the following definition \cite{kwek,nagler}:

\beq\label{generalizedBell}\ket{B^{\psi}_{m^*,n}}_{A,B}={1\over\sqrt{2}}\sum_{k=
0}^{1}
(-)^{k.n}\ket{\psi^*
_k}_{A}\ket{\psi_{k+m}}_{B}\eeq

Now, Eve's clone (labelled by $E$) is assumed to ``mirror'' Bob's qubit,
and as it is shown in \cite{nagler} the
formalism is simplified if we consider that the ancilla ($M$) ``mirrors''
Alice's qubit, which motivates the
following definition:

$\ket{B^{\psi}_{m,n^*}}_{E,M}={1\over\sqrt{2}}\sum_{k=0}^{1}
(-)^{k.n}\ket{\psi_k
}_{E}\ket{\psi^*_{k+m}}_{M}$

Note that when the $\psi$ basis is real, we recover the usual definition:

$\ket{B^{\psi}_{m^*,n}}_{A,B}=\ket{B^{\psi}_{m,n^*}}_{A,B}={1\over\sqrt
{2}}\sum_
{k=0}^{1}
(-)^{k.n}\ket{\psi_k}_{A}\ket{\psi_{k+m}}_{B}=\ket{B^{\psi}_{m,n}}_{A,B
}$

Accordingly, we shall postulate in the following the (modified) Cerf
ansatz: the cloning state is assumed to take
the following form:

\beqa \ket{\Psi}^{\psi}_{A,B,E,M}
=
\sum_{m,n=1}^
{N-1}a_{m,n}\ket{B^{\psi}_{m^*,n}}_{A,B}\ket{B^{\psi}_{m,n^*}}_{E,M
}\label{ansatzcerf}\eeqa .

On the basis of these definitions, it is natural to define strict
covariance as follows:

Definition:

The (generalised) Cerf cloner is said to be strictly
covariant
in the
$\psi$ basis and in the $\tilde\psi$ basis if and only if \beqa
\ket{\Psi}^{\psi}_{A,B,E,M}
=
\sum_{m,n=1}^
{N-1}a_{m,n}\ket{B^{\psi}_{m^*,n}}_{A,B}\ket{B^{\psi}_{m,n^*}}_{E,M
}=
\ket{\Psi}^{\tilde\psi}_{A,B,E,M}
=
\sum_{m,n=1}^
{N-1}a_{m,n}\ket{B^{\tilde\psi}_{m^*,n}}_{A,B}\ket{B^{\tilde\psi}_{
m,n^*}}_{E,M}\eeqa

In \cite{nagler}, the following theorem is shown:

Theorem:

The (generalised) Cerf cloner is strictly covariant in the
$\psi$ basis and in the $\tilde\psi$ basis if and only if whenever
$_{A,B}\braket{B^{\psi}_{m^*,n}}{B^{\tilde\psi}_{m'^*,n'}}_{A,B}\not= 0$
then $a_{m,n}=a_{m',n'}$.

This theorem provides an operational approach in order to build strictly
covariant (generalised) Cerf states: it is
sufficient to compute the in-products between the generalised Bell states
evaluated in the different bases;
whenever this in-product differs from zero, the corresponding elements of
the $a_{m,n}$ matrix that defines the
cloning transformation are equal.

Note that according to the identity \ref{identite}, for any pair of bases
$\{\psi$, $\tilde \psi\}$,
$_{A,B}\braket{B^{\psi}_{0^*,0}}{B^{\tilde\psi}_{m',n'^*}}_{A,B}=
\delta_{m',0}\delta_{n',0}$
so that
$a_{0,0}$ is a free and independent parameter for all cloning machines.

\subsection{3 qubit phase-covariant cloner (FGGNP cloner).}

We shall now apply these results in order to derive the 3 qubit
phase-covariant cloner. This cloner is aimed at
providing to Eve a strategy for eavesdropping the signal sent by Alice to
Bob during the completion of the
BB84 protocol. During this protocol, Eve encrypts Bob's signal (a fresh
cryptographic key) either into a state of
the $X$ basis, or into a state of the $Y$ basis. The 3 qubit
phase-covariant cloner is thus necessarily covariant
in the
$X$ basis and in the
$Y$ basis. We fix the phases of the
$X$ and $Y$ basis states as follows:

\beqa
&& \ket{+}^{X}={1\over \sqrt 2}(\ket{0}^Z+\ket{1}^Z),\qquad \ket{-}^{X}={1\over
\sqrt 2}(\ket{0}^Z-\ket{1}^Z)\\
&& \ket{+}^{Y}={1\over \sqrt 2}(\ket{0}^Z+i\ket{1}^Z),\qquad \ket
{-}^{Y}={1\over
\sqrt 2}(\ket{0}^Z-i\ket{1}^Z)\eeqa

Henceforth, it is easy to show the following equalities:
\beqa
&&\ket{+^*}^{X}=\ket{+}^{X}, \qquad \ket{-^*}^{X}=\ket{-}^{X}\\
&&\ket{+^*}^{Y}=\ket{-}^{Y}, \qquad \ket{+^*}^{Y}=\ket{+}^{Y}\eeqa
and also
\beqa
&&\ket{0}^Z={1\over \sqrt 2}(\ket{+}^{X}+\ket{-}^{X}), \qquad \ket{1}^Z={1\over
\sqrt 2}(\ket{+}^{X}-\ket{-}^{X})\\
&&\ket{0}^Z={1\over \sqrt 2}(\ket{+}^{Y}+\ket{-}^{Y}), \qquad
\ket{1}^Z={(-i)\over
\sqrt 2}(\ket{+}^{Y}-\ket{-}^{Y})\\
&&\ket{0}^Z={1\over \sqrt 2}(\ket{+^*}^{Y}+\ket{-^*}^{Y}),
\qquad\ket{1}^Z={i\over \sqrt
2}(\ket{+^*}^{Y}-\ket{-^*}^{Y})\eeqa

 From now on we shall omit to write the conjugation marks ($*$) when we
work in the
$Z$ basis or in the $X$ basis
because, as all the states of those bases have real amplitudes relatively
to the reference
basis states ($Z$). In order to improve the clarity, we shall
also always use the notations + and - for
labelling the spin up and down states on the equator, and reserve the
notations 0 and 1 for the spins polarized
along the $Z$ direction because this axis plays a special role in the
whole
treatment. One can establish the
following equalities by direct computation:

\beqa
\ket{B^{Z}_{0,0}}_{A,B}=\ket{B^{X}_{0,0}}_{A,B}=\ket{B^{Y}_{0^*,0}}_{A,B}=\ket{B
^{Y}_{0,0^*}}_{A,B}\nonumber\\
\ket{B^{Z}_{0,1}}_{A,B}=\ket{B^{X}_{1,0}}_{A,B}=\ket{B^{Y}_{1^*,0}}_{A,B}=\ket{B
^{Y}_{1,0^*}}_{A,B}\nonumber\\
\ket{B^{Z}_{1,0}}_{A,B}=\ket{B^{X}_{0,1}}_{A,B}=i\ket{B^{Y}_{1^*,1}}_{A,B}=(-i)\
ket{B^{Y}_{1,1^*}}_{A,B}\nonumber\\
\ket{B^{Z}_{1,1}}_{A,B}=-\ket{B^{X}_{1,1}}_{A,B}=(-i)\ket{B^{Y}_{0^*,1}}_{A,B}=i
\ket{B^{Y}_{0,1^*}}_{A,B}\label{X_Z}\eeqa

In virtue of the forementioned theorem, the state $\ket{\Psi}^{Z}_{A,B,E,M}$ is
strictly covariant in the $X$ and the $Y$
bases if and only if $a_{1,0}=a_{1,1}$. In the following, we shall
parametrise the matrix
$a_{m,n}$ of the phase-covariant cloner as follows: $a_{0,0}=v,
a_{0,1}=y, a_{1,0}=a_{1,1}=x$.

Then, in virtue of Eq.~(\ref{ansatz}), the cloning state of the
phase-covariant cloner obeys the following equation:
\beqa \label{covstate}\ket{\Psi}^{cov.}_{A,B,E,M}=
v\ket{B^{Z}_{0,0}}_{A,B}\ket{B^{Z}_{0,0}}_{E,M}
+y\ket{B^{Z}_{0,1}}_{A,B}\ket{B^{Z}_{0,1}}_{E,M}
+x(\ket{B^{Z}_{1,0}}_{A,B}\ket{B^{Z}_{1,0}}_{E,M}
+\ket{B^{Z}_{1,1}}_{A,B}\ket{B^{Z}_{1,1}}_{E,M})\eeqa

It is shown in \cite{kwek} that the
information gained by Eve is optimal when the
phases of the amplitudes $a_{m,n}$ are all the same. Physically, this
condition can be shown to ensure that
constructive interferences occur in certain detectors of Eve, and
destructive interference in others, in such a
way that Eve maximizes her information over Alice's key. Consistently, we
shall thus systematically assume in what
follows that the matrix $(a_{m,n})$ is a real matrix, with real positive
coefficients. Normalisation
of the phase-covariant cloner imposes
that 2$x^2$+$y^2$+$v^2$=1.

We can always define an ``equatorial'' basis that makes an angle $\phi$
with the $X$ basis on the Bloch
sphere through the expression
$\ket{+}^{\phi}={1\over
\sqrt 2}(\ket{0}^Z+e^{i\phi}\ket{1}^Z)$, $\ket{-}^{\phi}={1\over \sqrt
2}(\ket{0}^Z-e^{i\phi}\ket{1}^Z)$. By a
straightforward computation, it is easy to check that the following
relations are
valid:
\beqa &&\ket{B^{Z}_{0,0}}_{A,B}=
\ket{B^{\phi}_{0^*,0}}_{A,B}=\ket{B^{\phi}_{0,0^*}}_{A,B},\qquad
\ket{B^{Z}_{0,1}}_{A,B}=\ket{B^{\phi}_{1^*,0}}_{A,B}=\ket{B^{\phi}_{1,0^*}}_
{A,B}, \\
&&\ket{B^{Z}_{1,0}}_{A,B}=
cos \phi \ket{B^{\phi}_{0^*,1}}_{A,B}+isin \phi
\ket{B^{\phi}_{1^*,1}}_{A,B},\qquad \ket{B^{Z}_{1,1}}_{A,B}= (-)cos
\phi
\ket{B^{\phi}_{1^*,1}}_{A,B}-isin \phi \ket{B^{\phi}_{0^*,1}}_{A,B}\eeqa On
the
basis of these relations and on their
complex conjugates, it is easy to establish the strict covariance of the
cloner that satisfies $a_{1,0}=a_{1,1}$
on the whole equator. This justifies the name ``phase-covariant'' or
``equatorial'' cloner sometimes met in the
litterature.

  It is worth noting that, although in the seminal paper of FGGNP
\cite{FGGNP} the
generalized Bell states did not play any
special role, the phase-covariant cloner that they proposed can be shown
to
be equivalent to the phase-covariant
cloner derived in this section. Therefore, in the rest of the paper we
shall systematically refer to the phase-covariant
cloner as the 3 qubit FGGNP cloner: from now on, we shall denote the
phase-covariant state $\ket{\Psi}^{cov.}_{A,B,E,M}$
defined in Eq.~(\ref{covstate})  as $\ket{\Psi}^{FGGNP}_{A,B,E,M}$ and
consider that they are equivalent, although this
identification was made some years after the publication of FGGNP's paper.

Actually, if we consider a cloner that clones equally well the $Z$ and
the
$X$ bases,
then the constraint is
$a_{1,0}=a_{0,1}$, and the cloner that satisfies this constraint is
strictly covariant on the meridian that
passes through the poles and the $X$ basis states on the Bloch sphere
(the
big circle constituted by the
intersection of the $XZ$ plane and the Bloch sphere). We could say that
it
is a ``Greenwich-covariant'' cloner.
Moreover, as all the states on this meridian are purely real, the
generalized Cerf formalism reduces to the Cerf
formalism in this case. The ``Greenwich-covariant'' and the
phase-covariant
cloner obviously exhibit similar
properties. This explains why the equivalence between the
Greenwich-covariant cloner {\it a la Cerf} and the
phase-covariant FGGNP cloner was already established by N. Cerf in the past
(see for instance the appendix of reference
\cite{Cloning-a-qutrit}).

\subsection{2 qubit phase-covariant cloner (NG cloner), and connections
with the 3 qubit phase-covariant
cloner (FGGNP cloner).\label{comparison}}

As before, let us consider that Alice sends a fresh cryptographic key
according to the BB84 protocol, which
means that she encrypts the signal along a state $\ket{\psi}_{B}$ chosen
at
random among one of the four following
states:
$\ket{+}^X_{B}$,
$\ket{-}^X_{B}$, $\ket{+}^Y_{B}$ or $\ket{-}^Y_{B}$.

The NG copying machine \cite{NG} works as follows: Eve lets work on the
initial state
$\ket{\psi}_{B}\ket{0}^Z_{E}$ a unitary transformation of the type $
U_{BE}$
that is conceived in such a way that
$U_{BE}\ket{0}^Z_{B}\ket{0}^Z_{E}=\ket{0}^Z_{B}\ket{0}^Z_{E}$ and
$U_{BE}\ket{1}^Z_{B}\ket{0}^Z_{E}=cos\alpha \ket{1}^Z_{B}\ket{0}^Z_{E} +
sin\alpha \ket{0}^Z_{B}\ket{1}^Z_{E}$.

Remark that no ancilla ($M$) is required.

In order to implement BB84 protocol, Alice and Bob could as well share the
maximally entangled state
$\ket{B_{0,0}}_{A,B}$, while Alice would measure the spin of her qubit at
random
either along $X$ or along $Y$.
Then, the NG proposal can be formulated as follows: Eve
copies the state $\ket{B_{0,0}}^Z_{A,B}$ by replacing it by the cloning
state, which is assumed to be a 3-qubit
state (one qubit for Alice, one for Bob, one for Eve and no ancilla). We
shall from now on
denote this state $\ket{\Psi}^{NG}_{A,B,E}$:
\beq\ket{\Psi}^{NG}_{A,B,E}={1\over \sqrt
2}(\ket{0}^Z_{A} U_{BE}\ket{0}^Z_{B}\ket{0}^Z_{E}+\ket{1}^Z_{A}
U_{BE}\ket{1}^Z_{B}\ket{0}^Z_{E})={1\over \sqrt
2}(\ket{000}^Z_{ABE}+cos\alpha \ket{110}^Z_{ABE} +  sin\alpha
\ket{101}^Z_{ABE})\label{NG}\eeq.

By direct computation we obtain the following expressions for the NG
cloning state $\ket{\Psi}^{NG}_{A,B,E}$ in the
$X$ basis:

$\ket{\Psi}^{NG}_{A,B,E}={1\over 4}\{
(\ket{+}^X_{A}+\ket{-}^X_{A})(\ket{+}^X_{B}+\ket
{-}^X_{B})(\ket{+}^X_{E}+\ket{-}
^X_{E})+$

$cos\alpha
(\ket{+}^X_{A}-\ket{-}^X_{A})(\ket{+}^X_{B}-\ket
{-}^X_{B})(\ket{+}^X_{E}+\ket{-}
^X_{E})+$

$+
sin\alpha(\ket{+}^X_{A}-\ket{-}^X_{A})(\ket{+}^X_{B}+\ket{-}^X_{B})
(\ket{+}^X_{E}-\ket{-}^X_{E})\}$

In the $Y$ basis, we get:

$\ket{\Psi}^{NG}_{A,B,E}={1\over 4}\{
(\ket{+^*}^Y_{A}+\ket{-^*}^Y_{A})(\ket{+}^Y_{B}+\ket
{-}^Y_{B})(\ket{+}^Y_{E}+\ket{-}^Y_{E})+$

$cos\alpha
(i)(\ket{+^*}^Y_{A}-\ket{-^*}^Y_{A})(-i)(\ket{+}^Y_{B}-\ket
{-}^Y_{B})(\ket{+}^Y_
{E}+\ket{-}^Y_{E})+$

$+
sin\alpha(i)(\ket{+^*}^Y_{A}-\ket{-^*}^Y_{A})(-i)(\ket{+}^Y_{B}+
\ket{-}^Y_{B})(\ket{+}^Y_{E}-\ket{-}^Y_{E})\}$.

This proves the strict covariance of the NG cloning state in the $X$ and
$Y$ bases. It is legitimate to
investigate whether a connection could exist between the 2 qubit NG
cloning state defined in Eq.~(\ref{NG}) and the 3 qubit FGGNP
cloning state defined in Eq.~(\ref{covstate}).
Many
differences are manifest between this cloning state and the NG state: as
we
noted from the beginning, the FGGNP is
a 3 qubit state (actually three + 1 if we also take account of Alice's
qubit) while the NG state is a 2 qubit state
(actually two + 1), but we have also that the FGGNP cloning state is
symmetric under the spin flip along $Z$ (which exchanges $\ket{0}^Z$
and
$\ket{1}^Z$) or under the spin flips
along $X$ or $Y$, which is not true for the FGGNP cloning state. Actually,
this property is a very general property of the
Bell states that can be generalised to arbitrary
dimensions \cite{kwek,nagler}.

Another difference is that, according to Eq.~(\ref{NG}) the NG states form
a family of states that are
parametrized by one parameter only ($\alpha$), while, taking account of
the
normalisation, two parameters are
necessary for specifying the FGGNP state defined in Eq.~(\ref{covstate}).

Nevertheless, the optimal
state derived in \cite{FGGNP} and \cite{NG} exhibits the same properties:
the fidelity is then the same for
Alice
and Bob's clones and is equal to
${1\over 2}+{1\over
\sqrt 8}$. This corresponds to the parameters choices $\alpha={\pi\over
4}$, $v={1\over 2}+{1\over \sqrt
8},y={1\over 2}-{1\over \sqrt 8},$and
$x={1\over \sqrt 8}$.

In general (this is for instance the case with the universal cloning
machine \cite{bourenanne}), if Eve
suppresses the ancilla, she loses some useful and relevant information,
but
this is not true in the present
(optimal) case as is also shown in \cite{bourenanne}.

We shall now show that a subclass of the set of FGGNP states, for which the
ancilla can be dropped without losing information,
reduces to a one-parameter class of states
that ``looks like'' the NG state after elimination of the ancilla (the
qubit $M$). In order to do so, let us for instance consider that after
the
FGGNP cloning transformation, Alice
and Bob measure their respective qubit in the $X$ basis. Naturally, Eve,
who is assumed to listen to their public
communication gets informed about their choice of basis and decides to
measure her qubit and the ancilla in the $X$
basis too. Let us assume that Alice's measurement reveals that the state
of
the $A$ qubit is up ($\ket{+}^X_{A}$)
(as the full state is symmetric under the exchange of $\ket{+}^X$ and
$\ket{-}^X$, the treatment would be entirely
similar when she measures a spin down along $X$). Then, we obtain by
direct
computation that the probabilities
that Eve's qubit is polarised along $+X$ ($-X$) while the ancilla is
polarised along $+X$ ($-X$), denoted
$P_{EM}(+_X (-_X),+_X(-_X))$ are distributed as follows: $P_{EM}(+_X
,+_X)={1\over 2}(v+x)^2$,$P_{EM}(-_X
,-_X)={1\over 2}(v-x)^2$, $P_{EM}(+_X,-_X)={1\over 2}(y+x)^2$, and
$P_{EM}(-_X ,+_X)={1\over 2}(y-x)^2$. Now, as
it is shown in \cite{bourenanne,bruss6states}, the extra-information that is
present
in the ancilla is optimally exploited
if Eve conditions the result of the measurement on the qubit $E$ on its
equality\footnote{The forementioned
result \cite{kwek} according to which the information of Eve is
maximized, in arbitrary dimension, when the phases of the amplitudes
$a_{m,n}$ are real, concerns actually Eve's information {\bf conditioned}
on the {\bf difference} between results of measurements performed on the
clone $E$
 and results of measurements performed on the ancilla $M$).} (inequality)
with the result of
the
measurement on the ancilla $M$. We
shall now emit the heuristic hypothesis according to which it is possible
to drop the ancilla and thus to replace
the 3 qubit cloning state by a 2 qubit state when this can be done without
losing information so to say when the
statistics of Eve's results is invariant when we condition them on their
equality
(inequality)
with the values of the spin of the
ancilla. The validity of this hypothesis will be discussed in detail in
the
next section.

The ancilla can be dropped without losing
information whenever ${P_{EM}(+_x ,+_X)\over
P_{EM}(+_x ,-_X) }={P_{EM}(-_x ,-_X)\over P_{EM}(-_x ,+_X) }$. Under this
condition the probabilities of firing of
Eve's detectors are the same, when they are conditioned on their equality
(inequality) with the polarisations of
the ancilla. These constraints are fulfilled either when \beq{v+x\over x+y}
={v-x\over x-y }\eeq or when
\beq{v+x\over x+y} =-{v-x\over x-y }\eeq In addition of the normalisation
condition, each of these constraints defines a
one-parameter class of phase-covariant cloners, but the optimal cloner
fulfills only the first equation.
Therefore, from now on we shall focus only on the family defined by the
first equation. We shall elucidate the real meaning of this constraint
in the next subsection.
Taking account of the definition of the Bell states (Eq.~(\ref{Bell})), the
FGGNP state can be
rewritten as follows:

\beqa \ket{\Psi}^{FGGNP}_{A,B,E,M}= {1\over
2}\{v\ket{B^{Z}_{0,0}}_{A,B}\ket{0}^{Z}_{E}
+y\ket{B^{Z}_{0,1}}_{A,B}\ket{0}^{Z}_{E}
+x(\ket{B^{Z}_{1,0}}_{A,
B}-\ket{B^{Z}_{1,1}}_{A,B})\ket{1}^{Z}_{E}\}\ket{0}^{Z}_
{M}  +\nonumber\\
{1\over
2}\{v\ket{B^{Z}_{0,0}}_{A,B}\ket{1}^{Z}_{E}
-y\ket{B^{Z}_{0,1}}_{A,B}\ket{1}^{Z}_{E}
+x(\ket{B^{Z}_{1,0}}_{A,B}+\ket{B^{Z}_{1,1}}_{A,B})\ket{0}^{Z}_{E}\}\ket
{1}^{Z}_
{M}\eeqa

If ${v+x\over x+y} ={v-x\over x-y }$, then $x^2=vy$. Beside,
2$x^2$+$y^2$+$v^2$=1 by normalisation so that
$(v+y)^2=v^2+2vy+y^2=v^2+2x^2+y^2=1$. As $v$ and $y$ are real positive
parameters, we have that $v+y=1$. The
normalisation condition can also be rewritten as follows: $(v-
y)^2+4x^2=1$
so that we are free to redefine $v-y$ and $2x$
as follows: $v-y=cos \alpha$ and $2x=sin \alpha$, where $\alpha\in [0,\pi]$.
Taking account of these reparametrisations, and of the definition of Bell
states, we can finally express the FGGNP
cloning state as follows:
\beqa \ket{\Psi}^{FGGNP}_{A,B,E,M}= {1\over \sqrt
2}(\ket{000}^Z_{ABE}+cos\alpha \ket{110}^Z_{ABE} +  sin\alpha
\ket{101}^Z_{ABE})\ket{0}^{Z}_{M} +\nonumber \\ {1\over \sqrt
2}(\ket{111}^Z_{ABE}+cos\alpha \ket{001}^Z_{ABE} +
sin\alpha
\ket{010}^Z_{ABE})\ket{1}^{Z}_{M}\label{11}\eeqa
This means that
\beqa \label{identity} \ket{\Psi}^{FGGNP}_{A,B,E,M}= {1\over \sqrt
2}(\ket{\Psi}^{NG}_{A,B,E}\ket{0}^{Z}_{M}+\ket{\Psi}^{NGflip.}_{A,B,E}\ket{
1}^{Z
}_{M})\eeqa where
$\ket{\Psi}^{NGflip .}_{A,B,E}$ is obtained by inverting the north and
south poles (0 and 1) inside the
expression of $\ket{\Psi}^{NG}_{A,B,E}$. In other words, when we trace
over
the ancilla during the FGGNP cloning
process, everything happens as if we realised the NG attack with
probability 50 \% and the same attack, up to a
permutation of the north and the south pole, otherwise\footnote{One of us
(T.D) was informed by N. Cerf (private
communication), after the completion of this work, that he and N. Gisin
already derived a similar result in the past but
never published it.}. The FGGNP attack  is thus equivalent to a symmetrized
(relatively to the equatorial plane) version of the NG attack. For
instance, in the optimal case, the error rate
of the NG cloning process is equal to 50 \% along the south pole and to 0
along the north pole. In average it is
thus equal to 25 \% during the symmetrized NG process, the same as for
the
optimal FGGNP process, as it must.

\section{About the reducibility of 3 qubit cloners to 2 qubit
cloners.\label{lastsection}}

In the previous chapter, we showed that, when Eve does not gain more
information when
she conditions the results of the measurements performed on the clone onto
their equality with those obtained from the
ancilla, the 3 qubit cloner can be  reduced to a 2 qubit cloner. This
condition was introduced heuristically. It is  interesting to understand
why it is so, and also to
investigate the  generality of this condition: for instance it is
legitimate to understand whether or not it is a necessary
condition, or a sufficient  one.

Before doing so, it is worth recalling that in the present approach we are
only interested
in 3 qubit cloning states that fulfill Cerf's ansatz. This means that the
cloning state is biorthogonal in the Bell bases, which imposes that
the Eq.~(\ref{ansatz}) is fulfilled, where $a_{m,n}$ is a
(normalised) 2x2 matrix and where the Bell states are defined by
Eq.~(\ref{Bell}),
relatively to a reference basis, say the ($Z$) basis. The specification
of the amplitudes $a_{m,n}$
defines the cloning transformation.

The following definition that was inspired by the equality
~(\ref{identity}) helps us to
precise what we mean in general by reducibility of 3 qubits cloners
to 2 qubit cloners.

Main definition:

A 3 qubit cloner that fulfills Cerf's ansatz (in the reference ($Z$)
basis) is said to be reducible to a 2 qubit cloner in the $Z$ basis iff
it is possible to find a qubit basis $\{\ket{\tilde 0},\ket{\tilde
1}\}$, a positive real number $p$ comprised between 0 and 1 and two
unitary two-qubit transformations $U_{BE}$ and $V_{BE}$ such that
\beqa \ket{\Psi}^{Cerf}_{A,B,E,M} =\sum_{m,n=0}^1
a_{m,n}\ket{B_{m,n}}^Z_{A,B}\ket{B_{m,n}}^Z_{E,M}=\sqrt{p}\ket{\Psi^{U}}_{A,B,E}
\ket{\tilde 0}_{M}+
\sqrt{1-p}\ket{\Psi^{V}}_{A,B,E}\ket{\tilde 1}_{M}\label{reducible}\eeqa
where $\ket{\Psi^{U}}_{A,B,E}=1_A \otimes
U_{BE}\ket{B^Z_{0,0}}_{AB}\ket{0}^Z_{E}$
and $\ket{\Psi^{V}}_{A,B,E}=1_A \otimes
V_{BE}\ket{B^Z_{0,0}}_{AB}\ket{0}^Z_{E}$

It is very easy to check that the following corollaries are valid:

Corollary 1: a 3 qubit cloner is reducible to a qubit cloner iff
it is possible to find a qubit basis $\{\ket{\tilde 0},\ket{\tilde
1}\}$, and two unitary two-qubit transformations $U'_{BE}$ and $V'_{BE}$
such that
\beqa \ket{\Psi}^{Cerf}_{A,B,E,M}=
\sqrt{p}(\ket{\Psi^{U'}}_{A,B,E}\ket{\tilde 0}_{M}+\sqrt{1-p}
\ket{\Psi^{V'}}_{A,B,E}\ket{\tilde 1}_{M})\eeqa
where $\ket{\Psi^{U'}}_{A,B,E}=1_A \otimes
U'_{BE}\ket{B^{Z'}_{0,0}}_{AB}\ket{0}^{Z'}_{E}$
and $\ket{\Psi^{V'}}_{A,B,E}=1_A \otimes
V'_{BE}\ket{B^{Z'}_{0,0}}_{AB}\ket{0}^{Z'}_{E}$, where the $Z'$ basis is
arbitrary. This is a direct consequence of the identity \ref{identite}.

Corollary 2: when a 3 qubit cloner is reducible to a qubit cloner,
then the cloning state reduced on the degrees of freedom of the ancilla
is a mixture of two 2 qubit cloning states. Indeed, we have that
$\rho^{Cerf-
reduced}_{ABE}=Trace_{M}\ket{\Psi}^{Cerf}_{A,B,E,M}\bra{\Psi}^{Cerf}_{A,B,
E,M}=
p\ket{\Psi^{U}}_{A,B,E}\bra{\Psi^{U}}_{A,B,E}$+$(1-p)\ket{\Psi^{V}}_{A,B,E}\bra{
\Psi^{V}}_{A,B,E}.$ Moreover,
$\ket{\Psi^{U}}$ and $\ket{\Psi^{V}}$ can be generated by interactions that
do not involve Alice's qubit, and no ancilla $M$
is required, which represents a serious simplification of the cloning process.

We shall now prove the following theorem:

Main theorem:

A) Necessary condition:

When a 3 qubit cloner that fulfills Cerf's ansatz (in the reference
($Z$) basis) is reducible to a 2 qubit cloner in the $Z$ basis and when the
matrix $a_{m,n}$ is purely real,
then:

  either

i) $a_{0,0}a_{0,1}=a_{1,0}a_{1,1}$

or

ii) $a_{0,0}a_{1,0}=a_{0,1}a_{1,1}$

or

iii) $a_{0,0}a_{1,1}=a_{0,1}a_{1,0}$

B) Sufficient condition:

i) When $a_{0,0}a_{0,1}=a_{1,0}a_{1,1}$, then the Eq.~(\ref{reducible}) is
fulfilled with the qubit basis
$\{\ket{\tilde 0},\ket{\tilde  1}\}$ equal to the $Z$ basis.

ii) When $a_{0,0}a_{1,0}=a_{0,1}a_{1,1}$, then the Eq.~(\ref{reducible}) is
fulfilled with the qubit basis
$\{\ket{\tilde 0},\ket{\tilde  1}\}$ equal to the $X$ basis.

iii) When $a_{0,0}a_{1,1}=a_{0,1}a_{1,0}$, then the Eq.~(\ref{reducible})
is fulfilled with the qubit
basis
$\{\ket{\tilde 0},\ket{\tilde  1}\}$ equal to the $Y$ basis

{\bf Proof of the main theorem}:

A. Proof of the necessary condition.

 Without loss of
generality, we can parametrize $\{\ket{\tilde 0},\ket{\tilde  1}\}$ as
follows: $\{\ket{\tilde
0}=cos{\theta\over 2}\ket{0}+e^{i\phi}sin{\theta\over 2}\ket{1},
\ket{\tilde  1}=sin{\theta\over
2}\ket{0}-e^{i\phi}cos{\theta\over 2}\ket{1}\}$. Then, in virtue of
Eq.~(\ref{Bell}) and
Eq.~(\ref{reducible}), we have that
\beqa\ket{\Psi}^{Cerf}_{A,B,E,M} =\sum_{m,n=0}^1
a_{m,n}\ket{B_{m,n}}^Z_{A,B}\ket{B_{m,n}}^Z_{E,M}=\nonumber\\
{1\over
2}[\ket{0}^Z_A((a_{0,0}+a_{0,1})\ket{00}^Z_{BE}+(a_{1,0}-a_{1,1})\ket{11}^Z_{BE}
)\ket{0}^Z_M+\nonumber\\
\ket{0}^Z_A((a_{0,0}-a_{0,1})\ket{01}^Z_{BE}+(a_{1,0}+a_{1,1})\ket{10}^Z_{BE})\ket{1}^Z_M+\nonumber\\
\ket{1}^Z_A((a_{0,0}-a_{0,1})\ket{10}^Z_{BE}+(a_{1,0}+a_{1,1})\ket{01}^Z_{BE})\ket{0}^Z_M+\nonumber\\
\ket{1}^Z_A((a_{0,0}+a_{0,1})\ket{11}^Z_{BE}+(a_{1,0}-a_{1,1})\ket{00}^Z_{BE})\ket{1}^Z_M ]=\nonumber\\
{\sqrt{p}\over \sqrt
2}(\ket{0}^Z_A
U_{BE}\ket{00}^Z_{BE}+\ket{1}^Z_AU_{BE}\ket{10}^Z_{BE})(cos{\theta\over
2}\ket{0}^Z_M+e^{i\phi}sin{\theta\over
2}\ket{1}^Z_M)+\nonumber\\{\sqrt{1-p}\over
\sqrt 2}(\ket{0}^Z_A V_{BE}\ket{00}^Z_{BE}+\ket{1}^Z_A
V_{BE}\ket{10}^Z_{BE})(sin{\theta\over
2}\ket{0}^Z_M-e^{i\phi}cos{\theta\over 2}\ket{1}^Z_M)\label{equal}\eeqa

Unitarity of $U$ imposes that $U_{BE}\ket{00}^Z_{BE}$ and
$U_{BE}\ket{10}^Z_{BE}$ are mutually orthogonal
and normalised. A similar constraint holds for $V$. Now, projecting the
equality \ref{equal} onto the basis
$\ket{ij}_{AM}$ (where $i,j \in \{0,1\}$), we obtain four identities:
\beq ({1\over
2})((a_{0,0}+a_{0,1})\ket{00}^Z_{BE}+(a_{1,0}-a_{1,1})\ket{11}^Z_{BE})={\sqrt{p}
\over \sqrt
2}cos{\theta\over 2}U_{BE}\ket{00}_{BE}+{\sqrt{1-p}\over \sqrt
2}sin{\theta\over 2}V_{BE}\ket{00}_{BE}\eeq
\beq ({1\over
2})((a_{0,0}-a_{0,1})\ket{01}^Z_{BE}+(a_{1,0}+a_{1,1})\ket{10}^Z_{BE})={\sqrt{p}
\over \sqrt
2}e^{i\phi}sin{\theta\over 2}U_{BE}\ket{00}_{BE}+{\sqrt{1-p}\over \sqrt
2}(-)e^{i\phi}cos{\theta\over 2}V_{BE}\ket{00}_{BE}\eeq
\beq ({1\over
2})((a_{0,0}-a_{0,1})\ket{10}^Z_{BE}+(a_{1,0}+a_{1,1})\ket{01}^Z_{BE})={\sqrt{p}
\over \sqrt
2}cos{\theta\over 2}U_{BE}\ket{10}_{BE}+{\sqrt{1-p}\over \sqrt
2}sin{\theta\over 2}V_{BE}\ket{10}_{BE}\eeq
\beq ({1\over
2})((a_{0,0}+a_{0,1})\ket{11}^Z_{BE}+(a_{1,0}-a_{1,1})\ket{00}^Z_{BE})={\sqrt{p}
\over \sqrt
2}e^{i\phi}sin{\theta\over 2}U_{BE}\ket{10}_{BE}+{\sqrt{1-p}\over \sqrt
2}(-)e^{i\phi}cos{\theta\over 2}V_{BE}\ket{10}_{BE}\eeq
By elementary algebra, it is easy to check that:
\beq \sqrt{2p}e^{i\phi}U_{BE}\ket{00}^Z_{BE}=e^{i\phi}cos{\theta\over
2}((a_{0,0}+a_{0,1})\ket{00}^Z_{BE}+(a_{1,0}-a_{1,1})\ket{11}^Z_{BE})+sin{\theta
\over
2}((a_{0,0}-a_{0,1})\ket{01}^Z_{BE}+(a_{1,0}+a_{1,1})\ket{10}^Z_{BE})\eeq
\beq \sqrt{2(1-p)}e^{i\phi}V_{BE}\ket{00}^Z_{BE}=e^{i\phi}sin{\theta\over
2}((a_{0,0}+a_{0,1})\ket{00}^Z_{BE}+(a_{1,0}-a_{1,1})\ket{11}^Z_{BE})-cos{\theta
\over
2}((a_{0,0}-a_{0,1})\ket{01}^Z_{BE}+(a_{1,0}+a_{1,1})\ket{10}^Z_{BE})\eeq
\beq \sqrt{2p}e^{i\phi}U_{BE}\ket{10}^Z_{BE}=e^{i\phi}cos{\theta\over
2}((a_{0,0}-a_{0,1})\ket{10}^Z_{BE}+(a_{1,0}+a_{1,1})\ket{01}^Z_{BE})+sin{\theta
\over
2}((a_{0,0}+a_{0,1})\ket{11}^Z_{BE}+(a_{1,0}-a_{1,1})\ket{00}^Z_{BE})\eeq
\beq \sqrt{2(1-p)}e^{i\phi}V_{BE}\ket{10}^Z_{BE}=e^{i\phi}sin{\theta\over
2}((a_{0,0}-a_{0,1})\ket{10}^Z_{BE}+(a_{1,0}+a_{1,1})\ket{01}^Z_{BE})-cos{\theta
\over
2}((a_{0,0}+a_{0,1})\ket{11}^Z_{BE}+(a_{1,0}-a_{1,1})\ket{00}^Z_{BE})\eeq

 When $a_{m,n}$ is a purely real matrix, the orthogonality of
$U_{BE}\ket{00}_{BE}$ and
$U_{BE}\ket{10}_{BE}$ imposes that $e^{-i\phi}cos{\theta\over 2}sin{\theta\over
2}(a_{0,0}+a_{0,1})(a_{1,0}-a_{1,1})$+$e^{i\phi}cos{\theta\over
2}sin{\theta\over
2}(a_{0,0}-a_{0,1})(a_{1,0}+a_{1,1})=0$. This equation has two groups of
solutions: either $cos{\theta\over
2}sin{\theta\over 2}=0$, or
$(a_{0,0}+a_{0,1})(a_{1,0}-a_{1,1})=(-)e^{2i\phi}(a_{0,0}-a_{0,1})(a_{1,0}+a_{1,
1})$.

When $cos{\theta\over
2}sin{\theta\over 2}=0$, the basis $\{\ket{\tilde 0},\ket{\tilde  1}\}$ is
the $Z$ basis, and, because
$||U_{BE}\ket{00}_{BE}||$ = $||U_{BE}\ket{10}_{BE}||$ by unitarity, we get,
taking account of
the reality of the matrix $a_{m,n}$, that
$(a_{0,0}+a_{0,1})^2+(a_{1,0}-a_{1,1})^2=(a_{0,0}-a_{0,1})^2+(a_{1,0}+a_{1,1})^2
$, so that we must impose that (i)
$a_{0,0}a_{0,1}=a_{1,0}a_{1,1}$.

When
$(a_{0,0}+a_{0,1})(a_{1,0}-a_{1,1})=(-)e^{2i\phi}(a_{0,0}-a_{0,1})(a_{1,0}+a_{1,
1})$, taking account of
the reality of the matrix $a_{m,n}$, either $\phi$=0 and (ii)
$a_{0,0}a_{1,0}=a_{0,1}a_{1,1}$ or $\phi={\pi\over
2}$ and (iii) $a_{0,0}a_{1,1}=a_{0,1}a_{1,0}$.  It is easy to check that
when the conditions i, ii, or iii are
fulfilled the other constraints imposed by the unitarity of $U$ and $V$ are
automatically satisfied.

B. Proof of the sufficient condition.

Let us firstly prove the sufficient condition (ii). If
$a_{0,0}a_{1,0}=a_{0,1}a_{1,1}$, then
$(a_{0,0}+a_{0,1})(a_{1,0}-a_{1,1})=-(a_{0,0}-a_{0,1})(a_{1,0}+a_{1,1})$
and we can, without loss of generality, assume
that $(a_{0,0}+a_{0,1})=\alpha, (a_{0,0}-a_{0,1})=\beta, (a_{1,0}+a_{1,1})=
-r\alpha$, and $(a_{1,0}-a_{1,1})= r\beta$ with $r$,
$\alpha$ and $\beta$ real (in the special cases where $\alpha$ and $\beta$
would be equal to 0, we must be careful and
consider the limit in which
$r$ would go to infinity, but it does not invalidate the reasoning). We
have then, taking account of Eq.~(\ref{ansatz}) and
Eq.~(\ref{Bell}), after substitution, that

\beqa
\ket{\Psi}^{Cerf}_{A,B,E,M}=\nonumber\\
{1\over 2}\ket{0}^Z_A[ (\alpha\ket{00}^Z_{BE}+r\beta\ket{11}^Z_{BE})
\ket{0}^Z_M+(\beta\ket{01}^Z_{BE}-r\alpha\ket{10}^Z_{BE})\ket{1}^Z_M ]
+\nonumber\\
\ket{1}^Z_A[(\beta\ket{10}^Z_{BE}-r\alpha\ket{01}^Z_{BE})\ket{0}^Z_M+
(\alpha\ket{11}^Z_{BE}+r\beta\ket{00}^Z_{BE})\ket{1}^Z_M]=\nonumber\\
{1\over 2}
\ket{0}^Z_A(\alpha\ket{00}^Z_{BE}+r\beta\ket{11}^Z_{BE}+\beta\ket{01}^Z_{BE}-r\a
lpha\ket{10}^Z_{BE}){1\over \sqrt
2}(\ket{0}^Z_M+\ket{1}^Z_M)\nonumber\\
+\ket{1}^Z_A(\alpha\ket{11}^Z_{BE}+r\beta\ket{00}^Z_{BE}+\beta\ket{10}^Z_{BE}-r\
alpha\ket{01}^Z_{BE}){1\over \sqrt
2}(\ket{0}^Z_M+\ket{1}^Z_M)\nonumber\\
+\ket{0}^Z_A(\alpha\ket{00}^Z_{BE}+r\beta\ket{11}^Z_{BE}-\beta\ket{01}^Z_{BE}+r\
alpha\ket{10}^Z_{BE}){1\over \sqrt
2}(\ket{0}^Z_M-\ket{1}^Z_M)\nonumber\\
+\ket{1}^Z_A(-\alpha\ket{11}^Z_{BE}-r\beta\ket{00}^Z_{BE}+\beta\ket{10}^Z_{BE}-r
\alpha\ket{01}^Z_{BE}){1\over \sqrt
2}(\ket{0}^Z_M-\ket{1}^Z_M)\label{38}\eeqa

Let us consider the transformations $U$ and $B$ defined as follows:

\beqa
 U_{BE}\ket{00}_{BE}
={1\over \sqrt
2}(\alpha\ket{00}^Z_{BE}+r\beta\ket{11}^Z_{BE}+\beta\ket{01}^Z_{BE}-r\alpha\ket{
10}^Z_{BE})\nonumber\\
U_{BE}\ket{10}_{BE}
={1\over \sqrt
2}(\alpha\ket{11}^Z_{BE}+r\beta\ket{00}^Z_{BE}+\beta\ket{10}^Z_{BE}-r\alpha\ket{
01}^Z_{BE})\nonumber\\
V_{BE}\ket{00}_{BE}
={1\over \sqrt
2}(\alpha\ket{00}^Z_{BE}+r\beta\ket{11}^Z_{BE}-\beta\ket{01}^Z_{BE}+r\alpha\ket{
10}^Z_{BE})\nonumber\\
V_{BE}\ket{00}_{BE}
={1\over \sqrt
2}(-\alpha\ket{11}^Z_{BE}-r\beta\ket{00}^Z_{BE}+\beta\ket{10}^Z_{BE}-r\alpha\ket
{01}^Z_{BE})\nonumber\\
\label{39}\eeqa
It is easy to check that $U$ and $V$ are unitary, taking account of the
normalisation of $\ket{\Psi}^{Cerf}_{A,B,E,M}$ and of
the fact that $r$ is real. Substituting Eq.~(\ref{39}) into Eq.~(\ref{38})
ends the proof of the sufficient condition (ii).

The proof of the sufficient condition (iii) is entirely similar. For
proving the sufficient condition (i) it is enough to
make use of the identities \ref{X_Z}. Then, in virtue of the sufficient
condition (ii), we have that when
$a_{0,0}a_{0,1}=a_{1,0}a_{1,1}$,

\beqa \ket{\Psi}^{Cerf}_{A,B,E,M} =
{\sqrt{p}\over \sqrt
2}\ket{\Psi^{U}}_{A,B,E}\ket{\tilde 0}_{M}+{\sqrt{1-p}\over \sqrt
2}\ket{\Psi^{V}}_{A,B,E}\ket{\tilde 1}_{M}\eeqa where $\ket{\tilde
0}_{M}={1\over \sqrt
2}(\ket{+}^X+\ket{-}^X)=\ket{0}^Z_{M}$ and
$\ket{\tilde 1}_{M}={1\over \sqrt 2}(\ket{+}^X-\ket{-}^X)=\ket{1}_{M}^Z$,
and where $\ket{\Psi^{U}}_{A,B,E}=1_A \otimes
U_{BE}\ket{B^X_{0,0}}_{AB}\ket{+}^X_{E}=1_A \otimes
U'_{BE}\ket{B^X_{0,0}}_{AB}\ket{0}^Z_{E}$
and $\ket{\Psi^{V}}_{A,B,E}=1_A \otimes
V_{BE}\ket{B^Z_{0,0}}_{AB}\ket{+}^X_{E}=1_A \otimes
V'_{BE}\ket{B^Z_{0,0}}_{AB}\ket{0}^Z_{E}$ with $U,U',V,V'$ unitary, so that:

\beqa \ket{\Psi}^{Cerf}_{A,B,E,M}  ={\sqrt{p}\over \sqrt
2}\ket{\Psi^{U'}}_{A,B,E}\ket{0}^Z_{M}+{\sqrt{1-p}\over \sqrt
2}\ket{\Psi^{V'}}_{A,B,E}\ket{1}_{M}^Z\eeqa which ends the proof. Note that
in all the cases the symmetry of
$\ket{\Psi}^{Cerf}_{A,B,E,M}$ under permutation of the basis states of the
$X$, $Y$, and $Z$ bases imposes that $p$=$1-p$=
${1\over 2}$.

Let us now reconsider the heuristic hypothesis according to which it is
possible
to drop the ancilla and thus to replace
the 3 qubit cloning state by a 2 qubit state when the statistics of Eve's
results is invariant when we condition them on their equality
(inequality)
with the values of the spin of the
ancilla. According to this hypothesis, the ancilla can be dropped without
losing
information whenever ${P_{EM}(+_Z ,+_Z)\over
P_{EM}(+_Z ,-_Z) }={P_{EM}(-_Z ,-_Z)\over P_{EM}(-_Z ,+_Z) }$. When the
matrix $a_{m,n}$ is purely real, this
means that either $a_{0,0}a_{1,0}=a_{0,1}a_{1,1}$ or
$a_{0,0}a_{1,1}=a_{0,1}a_{1,0}$, which corresponds to the conditions $ii$
and
$iii$. Obviously the generality of this hypothesis is envalidated by the
sufficient condition $i$.

Note that the class of cloning
machines that was considered in the section \ref{comparison} corresponds to
the condition $ii$, excepted that the reference
basis is then the $X$ basis (or the $Y$ basis in virtue of the covariance).
The condition $ii$ must be rewritten after
permutation of $a_{1,0}$ and $a_{0,1}$, which is equivalent to the
condition $i$ in terms of the $Z$ basis, in
accordance with the identities \ref{X_Z}.

On the basis of the main theorem, it is easy to prove that the (optimal)
universal cloner is not reducible
to a 2 qubit cloner. Indeed, for such a cloner, $a_{0,1}=a_{1,0}=a_{1,1}$
and $a_{0,0}\not= a_{0,1}$ so that none of the
conditions
$i$
$ii$
$iii$ can be fulfilled.
\section{Conclusions and comments.}In summary, we analyzed the possibility
to reduce 3 qubit cloning to 2 qubit
cloning. From the point of view of practical realisability, it is
advantageous to economize one qubit in the cloning process.
Therefore it is interesting to know precisely when this reduction is
possible, and this was the goal of the main theorem
presented in the section \ref{lastsection}.

It
is worth noting that there are two distinct cases of reducibility that
enter into the scope of this theorem: when the
condition $ii$ or $iii$ is fulfilled, it is equivalent for Eve to replace
the 3 qubit cloning state by its 2 qubit
version, because she does not lose any information by doing so, but this is
not true when the condition $i$ is fulfilled.

Beside, the extent of validity of our main theorem is per se limited: it
could happen that a 3 qubit cloning machine can
be reduced to a mixture of more than two 2 qubit cloning machines, and it
is not possible, on the basis of our main theorem
to determine wheter or not this is the case. It is also limited to qubits,
and can be considered as a first step towards a
theory of economic qu$N$it cloners.

Moreover, it could happen that 3 qubit cloning states are well
approximated by mixtures of 2 qubit cloning states, and our main theorem is
silent about this situation. It could also
happen that mixtures of 2 qubit cloners provide a good approximation of a
non-reducible cloner along certain directions of
the Hilbert space only. For instance a mixture of 3 phase covariant
cloners, each of which being covariant along a great
circle orthogonal to one of the three directions $X$, $Y$, and $Z$,
exhibits the same fidelity in the $X$, $Y$ and $Z$ bases.
It provides a good candidate for cloning the cryptographic key exchanged
between Alice and Bob during the 6 states protocol in
which each state of encryption is chosen at random in one of these bases
\cite{bruss6states}. It is not as performant as the
symmetric universal qubit cloner of Buzek and Hillery, for which a fidelity
of 5/6 $\approx 0,8333\%$ is achieved
\cite{buzek}, and constitutes a less dangerous attack than the optimal
(asymmetric) isotropic cloner of threshhold fidelity
$\approx 0,8436$ described in \cite{bruss6states,bechmann,bourenanne}.
Nevertheless, it provides in principle a fidelity
equal to 2/3.$({1\over 2}+{1\over
\sqrt 8})$+$1/3.3/4\approx 0,8190 \% $, and it can be realised, in virtue
of the results of the second section during an
attack that consists of a mixture of six 2 qubit cloning processes, it is
thus reducible to a 2 qubit attack. Remark that the
fidelity of this attack is not the same along all directions on the Bloch
sphere but is maximal along the three canonical
bases.

Finally, one could object that the main theorem concerns only cloning
states that are pure and fulfill
Eq.~(\ref{ansatzcerf}), with purely real amplitudes $a_{m,n}$ but it can be
checked that all the interesting cloners
considered in the litterature are of this type. A partial elucidation of
why it is so can be found in \cite{kwek}.

\begin{acknowledgments}
T.D. is a Postdoctoral Fellow of the Fonds voor Wetenschappelijke Onderzoek-Vlaanderen.
This research was supported by the Belgian Office for Scientific, Technical
and Cultural Affairs in the framework of the Inter-University Attraction Pole
Program of the Belgian
governement under grant V-18., the Fund for Scientific Research - Flanders
(FWO-V), the
Concerted Research
Action ``Photonics in Computing'' and  the research council (OZR) of the
VUB. J.D. acknowledges the support by the Nature Science Foundation of China (Grant No. 10075041), the
National Fundamental Research Program (Grant No. 2001CB309300), and the ASTAR
Grant No. 012-104-0040.
\end{acknowledgments}

\end{document}